\documentclass[fleqn]{article}
\usepackage{amsfonts}

\catcode`\@=11 \@addtoreset{equation}{section}\catcode`\@=12

 \def\beq#1{\begin{equation}\label{#1}}
 \def\eeq{\end{equation}}

 \def\bear#1{\begin{eqnarray}\label{#1}}
 \def\ear{\end{eqnarray}}

 \def\nn{\nonumber}
 \def\sign{\mathop{\rm sign}\nolimits}

  \newcommand{\fnm}{\footnotemark}
 \newcommand{\fnt}{\footnotetext}

 \newcommand{\N}{ {\mathbb N} }
 \newcommand{\R}{ {\mathbb R} }

\newcommand{\eps}{\varepsilon}
\newcommand{\tri}{\triangle}
\newcommand{\p}{\partial}

\begin{document}

  \begin{center}

  \large \bf Black brane solutions governed by fluxbrane
  polynomials
  \end{center}

 \vspace{0.3truecm}

 \begin{center}

  \normalsize\bf V. D. Ivashchuk\fnm[1]\fnt[1]{e-mail:
  ivashchuk@mail.ru}

\vspace{0.3truecm}

 \it Center for Gravitation and Fundamental Metrology,
 VNIIMS, 46 Ozyornaya ul., Moscow 119361, Russia

 \it Institute of Gravitation and Cosmology,
 Peoples' Friendship University of Russia,
 6 Miklukho-Maklaya ul., Moscow 117198, Russia

\end{center}

\large

\begin{abstract}

 A family of composite black brane solutions in the
model with scalar fields and fields of forms is presented.
 The metric of any solution  is defined on a  manifold which contains a product of
 several  Ricci-flat ``internal'' spaces. The solutions  are governed by
moduli functions $H_s$  ($s = 1, ..., m$) obeying non-linear
differential equations with certain boundary conditions imposed.
These master equations are equivalent to Toda-like equations and
depend upon the non-degenerate ($m \times  m$)  matrix  $A$.  It
was conjectured earlier that the functions $H_s$  should be
polynomials if $A$ is a Cartan matrix for some semisimple
finite-dimensional Lie algebra (of rank $m$). It is shown  that
the solutions to master equations may be found by using so-called
fluxbrane polynomials  which can be calculated (in principle) for
any semisimple finite-dimensional Lie algebra. Examples of
dilatonic charged black hole ($0$-brane) solutions
 related to Lie algebras $A_1$, $A_2$, $C_2$ and  $G_2$ are considered.

\end{abstract}



\section{Introduction}

In this paper we deal with spherically-symmetric solutions with
horizon  defined on product manifolds containing several
Ricci-flat factor-spaces (with diverse signatures and dimensions).
Solutions of such type appear either in models with antisymmetric
forms and scalar fields \cite{BIM}-\cite{IMtop} or in models with
multi-component anisotropic fluid (MCAF) \cite{IMS1}-\cite{I-10}.
For black brane solutions with $1$-dimensional factor-spaces (of
Euclidean signatures) see \cite{CT,AIV,Oh} and references therein.

These  and  more general brane cosmological and spherically
symmetric solutions were obtained by reduction of the field
equations to the Lagrange equations corresponding to Toda-like
systems \cite{IMJ,IK}.

Here we consider black brane solutions  in the model with scalar
field and fields of forms, when certain relations on parameters
are imposed. The solutions are governed by  a set of functions
$H_s$, $s = 1, ..., m$, obeying non-linear differential equations
with certain boundary conditions imposed. These equations depend
upon the non-degenerate $m \times  m$  matrix  $A$.  It was
conjectured in \cite{IMp1} that the  moduli functions $H_s$ should
be polynomials when $A$ is a  Cartan matrix for some semisimple
finite-dimensional Lie algebra  ${\cal G}$  of rank $m$. In this
case we deal with special solutions to open Toda chain equations
related to the Lie algebra ${\cal G}$ \cite{T,B,OP,K} which are
integrable in quadratures.  The conjecture  from \cite{IMp1} was
verified  for the Lie algebras
 $A_m$, $C_{m+1}$, $m \geq 1$ in  \cite{IMp2,IMp3} by using the solutions
 to Toda chain equations corresponding to the  Lie algebras $A_m$ from
  \cite{And}.

Here we show that the black brane  solutions under consideration
may be also found by using so-called fluxbrane polynomials
\cite{Iflux}, which may be calculated (in principle)   for any
(semi)simple finite-dimensional Lie algebra using MATHEMATICA
\cite{Iv-02} or MAPLE \cite{GoIv-08,GoIv-09}.  For any solution we
find the Hawking temperature   as a function of $p_i$-parameters
of fluxbrane polynomials. Recently, in \cite{LY} a similar
approach appeared in a context of special Toda charged black hole
solutions corresponding to Lie algebras $A_m$, where a formal
relation for $A_m$ fluxbrane polynomials was obtained in a way
similar to our earlier consideration \cite{IMp2,IMp3} based on the
Anderson solution \cite{And}. Here we illustrate the general
approach by applying it to dilatonic charged black hole solutions
related to Lie algebras $A_1$, $A_2$, $C_2$ and  $G_2$.


\section{Black brane solutions}

We start with a  model governed by the action

 \beq{1a.1}
 S=\int d^Dx \sqrt{|g|}\bigl\{R[g]-h_{\alpha\beta}g^{MN}\p_M\varphi^\alpha
 \p_N\varphi^\beta-\sum_{a\in\tri}\frac{\theta_a}{n_a!}
 \exp[2\lambda_a(\varphi)](F^a)^2\bigr\},
 \eeq
 where $g=g_{MN}(x)dx^M\otimes dx^N$ is a metric,
 $\varphi=(\varphi^\alpha)\in\R^l$ is a vector of scalar fields,
 $(h_{\alpha\beta})$ is a  constant symmetric non-degenerate
 $l\times l$ matrix $(l\in \N)$, $\theta_a=\pm1$,

 \beq{1a.2a}
  F^a =    dA^a
  =  \frac{1}{n_a!} F^a_{M_1 \ldots M_{n_a}}
  dz^{M_1} \wedge \ldots \wedge dz^{M_{n_a}}
 \eeq
 is a $n_a$-form ($n_a\ge1$), $\lambda_a$ is a 1-form on $\R^l$:
 $\lambda_a(\varphi)=\lambda_{a \alpha }\varphi^\alpha$,
$a\in\tri$, $\alpha=1,\dots,l$. In (\ref{1a.1}) we denote $|g| =
|\det (g_{MN})|$, $(F^a)^2_g  = F^a_{M_1 \ldots M_{n_a}} F^a_{N_1
\ldots N_{n_a}} g^{M_1 N_1} \ldots g^{M_{n_a} N_{n_a}}$, $a \in
\tri$. Here $\tri$ is some finite set. In the models with one time
all $\theta_a = 1$ when the signature of the metric is $(-1,+1,
\ldots, +1)$.

In \cite{IMp1,IMp2,IMp3} we have obtained a  family of black brane
solutions to the field equations corresponding to the action
(\ref{1a.1}). These solutions are defined on the manifold

 \beq{1a.2}
  M = (R_{0}, + \infty) \times (M_0 = S^{d_0}) \times
 (M_1 = \R) \times \ldots \times M_n,
 \eeq
 and have the following form

\bear{2a.30}
 g= \Bigl(\prod_{s \in S} H_s^{2 h_s d(I_s)/(D-2)}
 \Bigr) \biggl\{ f^{-1} dR \otimes dR + R^2  g^0  \\
 \nn - \Bigl(\prod_{s \in S} H_s^{-2 h_s} \Bigr) f  dt \otimes dt +
 \sum_{i = 2}^{n} \Bigl(\prod_{s\in S}
  H_s^{-2 h_s \delta_{iI_s}} \Bigr) g^i  \biggr\},
 \\  \label{2a.31}
 \exp(\varphi^\alpha)= \prod_{s\in S} H_s^{h_s \chi_s
 \lambda_{a_s}^\alpha},
 \\  \label{2.32a}
 F^a= \sum_{s \in S} \delta^a_{a_s} {\cal F}^{s}, \ear
 where $f =1 - 2\mu/R^{d}$,

 \beq{2a.32}
{\cal F}^s= - \frac{Q_s}{R^{d_0}} \left( \prod_{s' \in
 S}  H_{s'}^{- A_{s s'}} \right) dR \wedge \tau(I_s), \quad s\in
 S_e, \eeq

\beq{2a.33} {\cal F}^s= Q_s \tau(\bar I_s), \quad s\in S_m. \eeq

Here $Q_s \neq 0$, $s\in S$, are charge densities, $R >R_0$, $R_0
 = (2 \mu)^{1/d} > 0$ and $ d = d_0 -1$.

 In  (\ref{2a.30}) $g^0$ is the canonical
metric on the unit sphere $M_0 =S^{d_0}$ and $g^i$ is a Ricci-flat
metric on $M_{i}$, $i=  2,\ldots,n$ and $\delta_{iI}= \sum_{j\in
I} \delta_{ij}$ is the indicator of $i$ belonging to $I$:
$\delta_{iI}=  1$ for $i\in I$ and $\delta_{iI}= 0$ otherwise. We
also denote   $g^1 = -dt \otimes dt$.

The  brane  set  $S$ is by definition
 \beq{1a.6} S=  S_e \cup S_m, \quad S_v=
 \cup_{a\in\tri}\{a\}\times\{v\}\times\Omega_{a,v},
 \eeq

 $v=  e,m$ and $\Omega_{a,e}, \Omega_{a,m} \subset \Omega$, where
 $\Omega = \Omega(n)$  is the set of all non-empty subsets of $\{
1, \ldots, n \}$, i.e. all branes do not ``live'' in $M_0$.

Any brane index $s \in S$ has the form $s =   (a_s,v_s, I_s)$,
where $a_s \in \tri$, $v_s =  e,m$ and $I_s \in \Omega_{a_s,v_s}$.
The sets $S_e$ and $S_m$ define electric and magnetic branes
correspondingly. In (\ref{2a.31}) $\chi_s  =   +1, -1$ for $s \in
 S_e, S_m$, respectively. All branes contain the time manifold $M_1
 = \R$, i.e.

 \beq{1a.7a}
 1 \in I_s, \qquad \forall s \in S.
 \eeq

All  manifolds $M_{i}$, $i > 0$, are assumed to be oriented and
connected and  the volume $d_i$-forms

\beq{1a.12} \tau_i  \equiv \sqrt{|g^i(y_i)|} \ dy_i^{1} \wedge
\ldots \wedge dy_i^{d_i}, \eeq

and signature parameters

\beq{1e.12} \varepsilon(i)  \equiv {\rm sign}( \det (g^i_{m_i
             n_i})) = \pm 1 \eeq

are well-defined for all $i= 1,\ldots,n$. Here $d_{i} =   {\rm
dim} M_{i}$, $i =   0, \ldots, n$, $d_0 > 1$, $d_1 = 1$ and for
any
 $I =   \{ i_1, \ldots, i_k \} \in \Omega$, $i_1 < \ldots < i_k$,
we denote

\beq{1a.13}
 \tau(I) \equiv \tau_{i_1}  \wedge \ldots \wedge
 \tau_{i_k}, \qquad d(I)  =  \sum_{i \in I} d_i,
 \quad   \eps(I) \equiv \eps(i_1) \ldots \eps(i_k).
 \eeq

Forms ${\cal F}^s$  correspond to electric and magnetic branes for
 $s\in S_e, S_m$, respectively. In (\ref{2a.33})
 $\bar I = \{0,\ldots,n\}\setminus I$ and $\tau (\bar I) =
   \tau_0 \wedge \tau(\{1,\ldots,n\}\setminus I)$, where
 $\tau_0$ is the volume form on $M_0 =S^{d_0}$.

The parameters  $h_s$ appearing in the solution satisfy the
relations: $h_s = (B_{s s})^{-1}$, where

\bear{1a.17}
 B_{ss'} = d(I_s\cap I_{s'})+\frac{d(I_s)d(I_{s'})}{2-D} +
 \chi_s\chi_{s'}\lambda_{a_s \alpha }\lambda_{ a_{s'} \beta}
 h^{\alpha\beta},
 \ear

 $s, s' \in S$, with $(h^{\alpha\beta})=(h_{\alpha\beta})^{-1}$ and
 $D =   1 + \sum_{i =   1}^{n} d_{i}$.  In (\ref{2a.31})
 $\lambda_{a_s}^{\alpha } =   h^{\alpha \beta} \lambda_{a_s \beta }$.

    The parameters  $B_{ss'}$
 are scalar products of certain $U^s$-vectors belonging to $\R^{n+l+1}$:
  $B_{ss'} = (U^s,U^{s'})$ \cite{IMJ,IMtop}.

 Here we assume that
 \bear{1a.B1}
  \qquad  \qquad ({\bf i}) \  K_s = B_{ss} \neq 0, \quad \forall s \in
  S,
  \\ \label{1a.B2}
  ({\bf ii}) \ {\rm det}(B_{s s'}) \neq 0,
  \ear
  i.e. the matrix $(B_{ss'})$ is a non-degenerate one.

We consider the matrix
 \beq{1a.18}
 (A_{ss'}) = \left( 2 B_{s s'}/B_{s' s'} \right),
 \eeq
$s, s' \in S$. Here  some ordering in $S$ is assumed.

 We denote
 \bear{1.eps1}
 \eps_s= \eps(I_s) \theta_{a_s} \ \ {\rm for} \ \ v_s = e, \\
 \label{1.eps2}
 \eps_s= -\eps[g] \eps(I_s) \theta_{a_s}, \ \ {\rm for} \ \ v_s = m,
 \ear
 $s\in S$,  $\eps[g] = \sign\det(g_{MN})$.

Functions $H_s > 0$ obey the equations

 \beq{2.2.1}
  R^{d_0} \frac{d}{dR} \left[ \left(1 -
  \frac{2\mu}{R^{d}}\right) \frac{ R^{d_0} }{H_s} \frac{d
  H_s}{dR} \right] =  B_s \prod_{s' \in S}  H_{s'}^{- A_{s s'}},
 \eeq
 with $B_s = \eps_s B_{s s} Q_s^2$
 and the boundary conditions imposed:
  \beq{2.2.1a}
   H_s |_{R =R_0 + 0} = H_{s0} \in (0, +\infty),
   \eeq
  and
  \beq{2.2.1b}
  H_s|_{R = +\infty} = 1,
  \eeq
  $s \in S$.

  Here we also impose  the following  condition
   \beq{2.2.1c}
     H_s   \ {\rm is \ smooth \ in} \ (R_{\epsilon}, +
     \infty),
   \eeq
  $s \in S$, where $R_{\epsilon}= (2 \mu)^{1/d} e^{- \epsilon}$,
  $\epsilon > 0$.  Then  the metric  has a regular horizon
  at $R^{d} =   2 \mu$ and has an asymptotically flat
  $(2 + d_0)$-dimensional section.

Due to (\ref{2a.32}) and  (\ref{2a.33}), the dimension of brane
worldvolume is defined by relations $d(I_s)= n_{a_s}-1$
for $s \in S_e$ and $d(I_s)= D- n_{a_s} -1$ for $S_m$. For a
$p$-brane: $p = p_s =   d(I_s)-1$.

{\bf  Restrictions on brane intersections.}

The composite black brane solutions under consideration take place if two
 restrictions on the sets of branes are obeyed.
These restrictions guarantee  the block-diagonal form of the
energy-momentum tensor.

{\bf Restriction 1.} {\em For any colour index $a\in\tri$ and
electromagnetic index $v= e,m$

\beq{2.R1} {\bf (R1)} \quad d(I \cap J) \leq d(I)  - 2, \eeq for
any $I,J \in \Omega_{a,v}$, $I \neq J$ (here $d(I) = d(J)$).
 }

{\bf Restriction 2} {\em For any colour index $a \in\tri$, $I \in
\Omega_{a,e}$ (electric brane set) and $J \in \Omega_{a,m}$
(magnetic brane set) \beq{2.2.3a} {\bf (R2)} \quad d(I \cap J)
\neq 0. \eeq
 }

{\bf The Hawking temperature.} The Hawking temperature
corresponding to the solution reads
 \beq{2a.36}
 T_H=   \frac{d}{4 \pi (2 \mu)^{1/d}}
 \prod_{s \in S} H_{s0}^{-h_s}, \eeq
where $H_{s0}$ are defined in (\ref{2.2.1a}).

This solution describes a set of charged (by forms) overlapping
branes ``living'' on submanifolds of $M_1 \times \dots \times
M_n$.

 \section{Polynomial structure of $H_s$ for  semisimple Lie algebras}

\subsection{Black brane polynomials}

Now we deal  with solutions to second order non-linear
differential equations  (\ref{2.2.1}) which may be rewritten as
follows

\beq{3.3.1}
 \frac{d}{dz} \left( \frac{(1 - 2\mu z)}{H_s}
 \frac{d}{dz} H_s \right) = \bar B_s
 \prod_{l =1}^{m}  H_{l}^{- A_{s l}}, \eeq

 where $H_s(z) > 0$,
  $z = R^{-d} \in (0, (2\mu)^{-1})$  and $\bar B_s =
 B_s/ d^2 \neq 0$.

 Eqs. (\ref{2.2.1a}) and  (\ref{2.2.1b})
 read

 \bear{3.3.2a}
  H_{s}((2\mu)^{-1} -0) = H_{s0} \in (0, + \infty), \\
 \label{3.3.2b} H_{s}(+ 0) = 1, \ear
  $s = 1,..., m$. Here we identify for simplicity  $S$ with the set of  $m = |S|$ numbers:
  $S = \{1,..., m \}$.

    The condition  (\ref{2.2.1c}) reads as follows
     \beq{3.3.2c}
      H_s(z) > 0 \ {\rm is \ smooth \ in} \ (0, z_{\epsilon}),
    \eeq
   $s = 1,..., m$, where $z_{\epsilon}= (2 \mu)^{-1} e^{ \epsilon d}$,
   $\epsilon > 0$.

Equations  (\ref{3.3.1})  are equivalent to Toda-type equations
\cite{IMp2,IMp3}.

   It was conjectured in \cite{IMp1}
 that  equations (\ref{3.3.1})-(\ref{3.3.2b})
 have  polynomial solutions  when  $(A_{s l})$ is a  Cartan matrix for
 some  semisimple finite-dimensional Lie algebra $\cal G$ of rank
 $m$.  In this case we get

 \beq{3.3.12}
 H_{s}(z) = 1 + \sum_{k = 1}^{n_s} P_s^{(k)} z^k, \eeq

 where $P_s^{(k)}$ are constants, $k = 1,\ldots, n_s$;
 $P_s^{(n_s)} \neq 0$, and

 \beq{3.2.20}
  n_s =   2 \sum_{l = 1}^m  A^{s l}
  \eeq
 $s = 1,..., m$, are the components of twice the  dual Weyl
 vector in the basis of simple  co-roots \cite{FS}.
 Here $(A^{sl}) = (A_{sl})^{-1}$.

 This conjecture  was verified for ${\bf A_m}$ and ${\bf C_{m+1}}$
 series of Lie algebras in \cite{IMp2,IMp3}. In the extremal case ($\mu
 = + 0$) an analogue of this conjecture was suggested
(implicitly) in \cite{LMMP}.

{\bf  ${\bf A_1} \oplus \ldots \oplus {\bf A_1}$ -case.} The
simplest example occurs in the orthogonal case : $B_{sl} =
(U^s,U^{l})= 0$, for  $s \neq l$ \cite{BIM,IMJ} (see also
\cite{CT,AIV,Oh} and refs. therein). In this case $(A_{s l}) =
{\rm diag}(2,\ldots,2)$ is a Cartan matrix for the semisimple Lie
algebra ${\bf A_1} \oplus \ldots \oplus {\bf A_1}$ and

 \beq{3.3.5}
 H_{s}(z) = 1 + P_s z, \eeq
 with $P_s \neq 0$,  satisfying

 \beq{3.3.5a}
 P_s(P_s + 2\mu) = -\bar B_s = - \eps_s K_s Q_s^2/d^2,
 \eeq
 $s = 1,..., m$. When all $\eps_s K_s < 0$  there
 exists a unique set of numbers  $P_s > 0$ obeying (\ref{3.3.5a}).

{\bf  $A_2$-case.}
 For the Lie algebra $\cal G$ coinciding with  ${\bf A_2} = sl(3)$
  we get $n_1 = n_2 =2$ and

 \beq{3.4.1} H_{s} = 1 + P_s z + P_s^{(2)} z^{2},
 \eeq

 where $P_s=  P_s^{(1)}$ and $P_s^{(2)} \neq 0$ are constants, $s = 1,2$.

 It was found in \cite{IMp1} that for $P_1 +P_2 + 4\mu \neq 0$
 (e.g. when all $P_s >0 $) the following relations take place

 \bear{3.4.5}
 P_s^{(2)} = \frac{ P_s P_{s +1} (P_s + 2 \mu )}{2
(P_1 +P_2 + 4\mu)}, \qquad \bar B_s = - \frac{ P_s (P_s + 2 \mu
 )(P_s + 4 \mu )}{P_1 +P_2 + 4\mu}, \ear
 $s = 1,2$.

Here we denote $s+ 1 = 2, 1$ for $s = 1,2$, respectively.

 {\bf Other solutions.}
  The ``master'' equations were integrated (using
  Maple) in    \cite{GrIvKim1,GrIvMel2} for Lie
  algebras ${\bf C_2}$ and ${\bf A_3}$, respectively.
  Recently, in \cite{LY} the solutions to master equations were found for  $A_m$ Lie algebras
  using the general solutions to Toda chain equations from
  \cite{And}.

  Special solutions $H_{s}(z) = (1 + P_s z)^{b_s}$
   appeared earlier in  \cite{Br1} and later in \cite{IMJ2,CIM} in a context of
  so-called block-orthogonal configurations.

\section{Fluxbrane polynomials}

Here we deal with the so-called fluxbrane polynomials which will
be used in the next section for solving the black brane master
equations (\ref{3.3.1}) with the boundary conditions
(\ref{3.3.2a}) and  (\ref{3.3.2b}) imposed.

\subsection{The conjecture on fluxbrane polynomials}

Now,  we  deal with the polynomials  which obey the following set
of equations
  \beq{4.1}
  \frac{d}{dz} \left( \frac{z}{{\cal H}_s} \frac{d}{dz} {\cal H}_s \right) =
   {\cal P}_s  \prod_{l = 1}^{m}  {\cal H}_{l}^{- A_{s l}},
  \eeq
 with  the  boundary conditions imposed
 \beq{4.2}
   {\cal H}_{s}(+ 0) = 1,
 \eeq
 $s = 1,...,m$. Here  functions ${\cal H}_s(z) > 0$ are
 defined on the interval $(0, +\infty)$ if
 ${\cal P}_s > 0$  for all $s$ and $(A_{s l})$ is the Cartan matrix
 for some finite-dimensional  semisimple Lie algebra $\cal G$
 of rank $m$.

 The functions ${\cal H}_s(z) > 0$ appeared in
 generalized fluxbrane solutions which were obtained in  \cite{Iflux}.
 Parameter  ${\cal P}_s$ is proportional
 to brane charge density squared $q_s^2$,  $s = 1,...,m$  and $z = \rho^2$, where
  $\rho$ is a radial coordinate. The boundary condition
 (\ref{4.2}) guarantees the absence of singularity
 (in the metric) for $\rho =  +0$.
  For fluxbrane solutions in supergravities (or stringy-inspired models) see
  \cite{Iflux,CGS,GutSt,CGSaf} and references therein.

  The fluxbrane solutions from \cite{Iflux} and similar
  $S$-brane solutions from \cite{GonIM}  are special classes of
  more general solutions from \cite{IK}. The simplest
  ``fluxbrane'' solution  is a well-known Melvin solution
  \cite{Melv} corresponding to the Lie algebra $A_1 = sl(2)$.

 It was conjectured in \cite{Iflux} that eqs. (\ref{4.1}),
 (\ref{4.2})  have polynomial solutions
 \beq{4.3}
 {\cal H}_{s}(z) = 1 + \sum_{k = 1}^{n_s} {\cal P}_s^{(k)} z^k,
 \eeq
 where ${\cal P}_s^{(k)}$ are constants, $k = 1,\ldots, n_s$. Here
 ${\cal P}_s^{(n_s)} \neq 0$  and $n_s$ are defined in
 (\ref{3.2.20})

 It was pointed in \cite{Iflux} that the conjecture on polynomial
 structure of ${\cal H}_{s}$   may be
 verified for $A_n$ and $C_n$ Lie algebras along a line as it
 was done for black brane polynomials in \cite{IMp2,IMp3}.


The substitution of (\ref{4.3})  into (\ref{4.1}) gives a  chain
of relations on parameters ${\cal P}_s^{(k)}$  and
 ${\cal P}_s $. For $A-D-E$ (simply laced) Lie algebras these relations were used
 for calculations of polynomials (by using MATHEMATICA) in \cite{Iv-02}.  The first relation in this chain is
 \beq{4.5a}
  {\cal P}_s^{(1)} = {\cal P}_s ,
 \eeq
 $s = 1, ..., m$.

 We note that for a special choice
 of  parameters: ${\cal P}_s = n_s P$, $P > 0$, the polynomials
 have the following simple form \cite{Iflux}
 \beq{4.s}
 {\cal H}_{s}(z) = (1 +  P z)^{n_s},
 \eeq
 $s = 1, ..., m$. This ansatz is a nice
 tool for verification of general solutions obtained
 by either  analytical or computer calculations.

 In \cite{GoIv-08,GoIv-09}  a computational program for
calculations  of polynomials (using MAPLE) corresponding  to
classical series of simple Lie algebras was suggested.

  The calculations of  ${\cal P}_s^{(k)}$ (which are polynomials
  of $k$-th power in  ${\cal P}_s$) give huge denominators
  for big $k$.  This may be avoided by using new parameters $p_s$
  instead of ${\cal P}_s$
  \beq{4.g}
   p_s = {\cal P}_s/n_s,
  \eeq
  $s = 1, ..., m$.

\subsection{Examples of polynomials }

 Here we present certain examples of
 polynomials corresponding to the Lie algebras
 $A_2$,  $C_2$ and $G_2$.

 \subsubsection{$A_r$-polynomials, $r = 1,2,3$.}

  {\bf $A_1$-case.} The simplest example occurs in the case of
 the Lie algebra $A_1 = sl(2)$. Here $n_1 = 1$.
  We get

 \beq{A.1}
  {\cal H}_{1}(z) = 1 + p_1 z.
 \eeq

{\bf $A_2$-case.} For the Lie algebra $A_2 = sl(3)$ with the
Cartan matrix

 \beq{A.5}
    \left(A_{ss'}\right)=
  \left( \begin{array}{*{6}{c}}
     2 & -1\\
     -1& 2\\
    \end{array}
 \right)\quad \eeq
 we have \cite{Iflux}  $n_1 = n_2 =2$ and
 \bear{A.6}
  {\cal H}_{1} = 1 + 2 p_1 z +  p_1 p_2 z^{2}, \\
  \label{A.7}
  {\cal H}_{2} = 1 + 2 p_2 z +  p_1 p_2 z^{2}.
  \ear

 {\bf $A_3$-case.} The polynomials for the $A_3$-case read as
 follows

 \bear{A.8}
  {\cal H}_{1} = 1 +  3 p_1 z + 3 p_1 p_2 z^{2} +  p_1 p_2 p_3 z^{3}, \\
 \label{A.9}
    {\cal H}_{2} = 1 + 4 p_2 z + 3 (  p_1 p_2 +
      p_2 p_3 ) z^{2} + 4 p_1 p_2 p_3 z^{3}
    +  p_1 p_2^{2} p_3 z^{4},\\
 \label{A.10}
 {\cal H}_{3} = 1 + 3 p_3 z + 3 p_2 p_3 z^{2} +  p_1 p_2 p_3 z^{3}.
 \ear
  \vspace{15pt}

 \subsubsection{$C_2$-polynomials.}

For the Lie algebra $C_2 = so(5)$ with the Cartan matrix

 \beq{C.1}
    \left(A_{ss'}\right)=
  \left( \begin{array}{*{6}{c}}
     2 & -1\\
     -2& 2\\
 \end{array}
 \right)\quad
 \eeq
 we get from  (\ref{3.2.20})  $n_1 = 3$ and $n_2 = 4$.

 For fluxbrane polynomials we obtain from \cite{GonIM}
 \bear{C.2}
  {\cal H}_1= 1+ 3 p_1 z+ 3 p_1 p_2 z^2 + p_1^2 p_2 z^3,
               \\ \label{C.3}
  {\cal H}_2 = 1+ 4 p_2 z+ 6 p_1 p_2 z^2 + 4 p_1^2 p_2 z^3
        +  p_1^2 p_2^2 z^4.
 \ear

  \subsubsection{$G_2$-polynomials.}

For the Lie algebra $G_2$ with the Cartan matrix

\beq{G.1}
    \left(A_{ss'}\right)=
  \left( \begin{array}{*{6}{c}}
     2 & -1\\
     -3& 2\\
 \end{array}
 \right)\quad
 \eeq
 we get from  (\ref{3.2.20})  $n_1 = 6$ and $n_2 = 10$.
The fluxbrane polynomials read \cite{GonIM}
 \vspace{15pt}
\bear{G.2}
 {\cal H}_{1} = 1+ 6 p_1 z+ 15 p_1 p_2 z^2 + 20 p_1^2 p_2 z^3 + \\ \nonumber
    15 p_1^3 p_2 z^4 + 6 p_1^3 p_2^2 z^5 + p_1^4 p_2^2 z^6 ,
\\ \label{G.3}
  {\cal H}_2 =  1+  10 p_2 z + 45 p_1 p_2 z^2  +  120 p_1^2 p_2 z^3
  +  p_1^2 p_2 (135 p_1 + 75 p_2) z^4 \\ \nonumber
                + 252 p_1^3 p_2^2 z^5
   + p_1^3 p_2^2 \biggl(75 p_1 + 135 p_2 \biggr)z^6   +  120 p_1^4 p_2^3  z^7
   \\ \nonumber
   + 45 p_1^5 p_2^3 z^8 +  10 p_1^6 p_2^3 z^9
      + p_1^{6} p_2^{4}  z^{10}.
  \ear
 \vspace{15pt}

In all examples presented above the substitution $p_1 = \dots =
p_n = P$ into polynomials gives us  relations (\ref{4.s}).

In what follows we denote
  \beq{4.H}
  {\cal H}_{s}(z) = {\cal H}_{s}(z; p),
  \eeq
   $s = 1, ..., m$, where $p = (p_1,...,p_m)$.

 It should be noted that the set of fluxbrane polynomials for any semisimple Lie algebra
${\cal G} = {\cal G}_1 \oplus \dots \oplus {\cal G}_k$, where all
${\cal G}_i$ are simple Lie algebras, is just a union of sets of
fluxbrane polynomials, corresponding to ${\cal G}_1$, ... , ${\cal
G}_k$, respectively.

 \subsection{Toda chain solutions}

  Fluxbrane  polynomials ${\cal H}_s$ define  special solutions to  open Toda chain
  equations \cite{T,B,K,OP}
 \beq{4.T}
 \frac{d^2 q^s}{du^2} = -  4 n_s p_s \exp( \sum_{l = 1}^{m} A_{s l} q^{l}
 ),
 \eeq
which correspond to a semisimple finite dimensional  Lie algebra
$\cal G$ with the Cartan matrix $(A_{s l})$.

These solution read
 \beq{4.h}
   q^s(u) = - \ln {\cal H}_s(e^{-2u};p)  -  n_s u,
 \eeq
  $s = 1, ..., m$  (i.e. we put here $z = e^{-2u})$.

 Relations  (\ref{4.h})
 imply the following asymptotic formulae

 \bear{4.s1}
   q^s(u) =  - \ln C_s   +  n_s u + o(1),   \qquad {\rm for}
   \  u \to - \infty, \\ \label{4.s2}
    q^s(u) = - n_s p_s e^{-2u} (1 + o(1) )   -  n_s u,   \qquad {\rm for}
   \  u \to + \infty,
 \ear
 $s = 1, ..., m$. Here the scattering data $C_s > 0$  appear
 in the leading terms of  polynomials
  \beq{4.L}
 {\cal H}_s = 1 + n_s p_s z + \dots + C_s z^{n_s},
  \eeq
  $s = 1, ..., m$. The calculations (see examples above) give the following relations

   \beq{4.C}
   C_s  = \prod_{i=1}^{m} p_{i}^{\nu(s,i)},
    \eeq

   where ${\nu(s,i)}$ are non-negative integer numbers which obey the relations
  $\sum_{i=1}^m {\ \nu(s,i)} = n_s$, $s = 1, ..., m$.
   (They are natural ones for simple finite-dimensional Lie algebras.)

 It follows from (\ref{4.s1}) and (\ref{4.s2}) that

 \beq{4.v}
   v^s (\pm \infty) = \mp n_s, \qquad {\rm for}    \  u \to \pm
   \infty,
    \eeq
 $s = 1, ..., m$, i.e. the asymptotical values of velocities for ``in'' and ``out'' asymptotics
 are opposite in sign and coinciding by absolute value with
 the vector $(n_1, \dots, n_m)$.
 This is the main specific feature of the special solutions (\ref{4.h})
 to open Toda chain equations  which are described by fluxbrane polynomials.

\section{Black brane solutions governed by fluxbrane polynomials}

In this section we show that the ``black brane'' master equations
(\ref{3.3.1})-(\ref{3.3.2b}) may be solved (at least for small
enough values of charge parameters $Q_s$) in terms of fluxbrane
polynomials and give several examples of solutions.

\subsection{Reduction to  fluxbrane polynomials}

Let us denote $f =f(z)= 1 - 2\mu z$.
Then the relations
(\ref{3.3.1}) may be rewritten as \beq{5.1}
 \frac{d}{df} \left( \frac{f}{H_s}
 \frac{d}{df} H_s \right) =  B_s (2 \mu d )^{-2}
 \prod_{l =1}^{m}  H_{l}^{- A_{s l}}. \eeq
 $s = 1, ..., m$.
 These relations could be solved by using fluxbrane
polynomials ${\cal H}_{s}(f) = {\cal H}_{s}(f; p)$, corresponding
to $m \times m$ Cartan matrix $(A_{s l})$, where $p =
(p_1,...,p_m)$ is the set of reduced parameters (\ref{4.g}). Here
we impose the restrictions $p_i \neq 0$ for all $i$ instead of
$p_i > 0$ from the previous section.

We put
\beq{5.2}
  H_s(z) = {\cal H}_{s}(f(z);p)/{\cal H}_{s}(1;p)
\eeq for all  $s = 1, ..., m$. \fnm[2]\fnt[2]{This trick was known
to the author for many years as well as the schemes of
calculations of  ${\cal H}_{s}$ \cite{Iv-02}.}

Then the relations (\ref{5.1}), or, equivalently, (\ref{3.3.1})
are satisfied identically if

\beq{5.3}
 n_s p_s  \prod_{l =1}^{m}  ({\cal H}_{l}(1;p))^{- A_{s l}} = B_s /(2 \mu d)^2
 =  \eps_s K_{s} Q_s^2/(2 \mu d)^2,
\eeq $s = 1, ..., m$.

We call the set of parameters $p = (p_1,...,p_m)$ ($p_i \neq 0$)  as proper one if
  \beq{5.4}
 {\cal H}_{s}(f;p) > 0
 \eeq
 for all $f \in [0,1]$ and $s = 1,..., m$.

 We denote the set of all proper $p$ by ${\cal D}$.
 In what  follows we consider only proper $p$.

 Relations (\ref{5.3})  imply
\beq{5.5}
  {\rm sign} p_s   = {\rm sign} B_s = \eps_s,
\eeq
$s = 1, ..., m$.

The  boundary conditions  (\ref{3.3.2a}) are valid since
  \beq{5.6}
  H_{s}((2\mu)^{-1} -0) = 1/{\cal H}_{s}(1;p) > 0,
  \eeq
 $s = 1,..., m$,  and conditions (\ref{3.3.2b}) are satisfied just due to
 definition (\ref{5.2})  ($p \in {\cal D})$.

  Relations (\ref{5.6}) imply the following formula for
  the Hawking temperature  (\ref{2a.36})
 \beq{5.7}
 T_H=   \frac{d}{4 \pi (2 \mu)^{1/d}}
 \prod_{s = 1}^{m} ({\cal H}_{s}(1;p))^{h_s}, \eeq

It should be noted that usual black brane solutions deal with
negative $B_s$ (since $\eps_s < 0$ and $B_{ss} >0$) and hence
$p_s$ should be negative. Fluxbrane polynomials with negative
$p_s$ were considered earlier in cosmological ($S$-brane)
solutions which describe an accelerated expansion of 3-dimensional
factor-space \cite{IKM,Gol}. The original fluxbrane polynomials
\cite{Iflux} responsible for  fluxbrane solutions have positive
parameters $p_s$. For black brane applications positive $p_s$
correspond to positive $B_s$. This  takes place for a certain
family of phantom black brane solutions which may be a subject of
a separate publication.

  {\bf Remark.} For all examples of fluxbrane polynomials
  we know, all $p$ = $(p_i)$ with positive
  entries $p_i >0$ are proper since in this case ${\cal H}_{s}(f;p) >0$ for
  all $s$ and  $f \geq 0$.

 For  fixed signs in (\ref{5.5}) we denote
  \beq{5.8d}
  {\cal D}_{\epsilon} = \{p | p = (p_1, ...,p_m) \in {\cal D}, {\rm sign} p_i =
  \epsilon_i, i = 1,..., m \}.
  \eeq
 Here ${\epsilon} = (\epsilon_i)$.

 Then relations (\ref{5.3}) define
 the  map
 \beq{5.8}
  f_{\epsilon}: {\cal D}_{\epsilon} \to  \R_{+}^{m},
 \eeq
 with $f_{\epsilon}(p) = (Q_1^2, \dots, Q_m^2)$. Locally, for small enough
 $p_i$ the function $f_{\epsilon}$ defines one-to-one correspondence between
 the sets of parameters $(p_1, ...,p_m)$ and $(Q_1^2, \dots, Q_m^2)$.
 An open question here is to verify whether this is correct globally
 for certain choices of $\epsilon = (\epsilon_i)$ and semisimple Lie
 algebras ${\cal G}$ (i.e. whether the function
 $f_{\epsilon}$ is bijective or not for certain cases).

\subsection{Examples}

Now we illustrate the general approach by considering several
examples of charged black hole ($0$-brane) solutions corresponding
to Lie algebras of rank $m = 1, 2$.

 \subsubsection{Black hole for  $A_1$.}

First we consider the gravitational model with one scalar field
and one  2-form:
 \beq{6.1a}
  S=\int d^Dx \sqrt{|g|} \biggl \{R[g]-
  g^{MN} \partial_M \varphi \partial_N \varphi-
  \frac{1}{2}   \exp(2 \lambda \varphi)F^2
  \biggr \}.
 \eeq

Here $g$ is a D-dimensional metric, $F = dA$ is $2$-form;
$\varphi$ is scalar field and  $\lambda \in  \R$ is dilatonic
coupling. We deal with a charged black hole solution defined on
the manifold

 \beq{6.2}
  M = (0, + \infty) \times M_0 \times M_1 \times M_2,
 \eeq
 where $M_0 = S^{d_0}$, $M_1 = \R$ is a one-dimensional (time) manifold of signature $(-)$ and
 $M_2$ is  $d_2$-dimensional Ricci-flat manifold.

We put the brane multi-index  corresponding to 1-forms $A$  as
$I_1 = \{ 1 \}$. We get from (\ref{1a.B1})
   \beq{6.3b}
  K_1 =  \frac{D-3}{D-2} + \lambda^2 > 0.
  \eeq

In this case relation  (\ref{1a.18}) is valid for the Cartan
matrix $A = (2)$, corresponding to the Lie algebra $A_1$.

The charged  black hole solution has the following form (see
(\ref{2a.30})-(\ref{2a.32}))

 \bear{6.4b}
  \qquad g =  H^{2 h/(D-2)}
  \biggl\{ f^{-1} dR \otimes dR + R^2  g^0
    - H^{-2 h}  f  dt \otimes dt +
  g^2  \biggr\},  \\  \label{6.5b}
 \exp(\varphi)=  H^{h \lambda }
 \\  \label{6.6b}
 F =   \frac{Q}{R^{d_0}}  H^{- 2}   dt \wedge dR .
 \ear

 Here $f =1 - 2\mu/R^{d}$, $\mu > 0$, $d = d_0 -1$, $h = K_1^{-1} >0$,
 $Q = Q_1 \neq 0$  is (electric) charge, $R > R_0 = (2\mu)^{1/d}$,
 and $g^2$ is a Ricci-flat metric on $M_{2}$.

The moduli function $H$ is given by (\ref{A.1}) and general
prescriptions (\ref{5.2}) and (\ref{5.3}):

\beq{6.7b}
  H(z) =   \frac{1 + p_1 f(z)}{1 + p_1},
  \eeq

where    $f(z) = 1 - 2 \mu z$,  and the parameter $ p_1$ is negative due to relation

\bear{6.8b}
  p_1 / (1 + p_1)^2 =  -  K_1 Q^2 /(2 \mu d)^2,
 \ear
following from (\ref{5.3}).

 The parameter $p_1$ should be proper, i.e. ${\cal H}_{1}(f; p_1)  = 1 + p_1 f
> 0$ for all $f \in [0,1]$, hence we get $-1 < p_1 < 0$.
Function (\ref{5.8}) in this case is given by (\ref{6.8b})
 $f_{\epsilon}: {\cal D}_{\epsilon} = (-1,0) \to  \R_{+}$,
 $\epsilon = (-1)$. It is a bijection (moreover, $f_{\epsilon}$ is a diffeomorphism).

 Relation (\ref{6.7b}) may be rewritten as follows
 \beq{6.9bb}
  H(z) = 1 + P z, \qquad P = \frac{2 \mu (- p_1)}{1 + p_1} > 0.
 \eeq

 It follows from (\ref{6.8b}) and (\ref{6.9bb}) that

  \beq{6.10b}
  P(P + 2\mu) =   K_1 Q^2/d^2 = \frac{Q^2 A(\lambda)}{(D-2)d^2} , \qquad
  A(\lambda) = D-3 + \lambda^2 (D -2),
 \eeq
 in agreement with (\ref{3.3.5a}) and hence ($P >0$)
   \beq{6.11b}
  P = - \mu + \sqrt{\mu^2 + Q^2 A(\lambda)(D-2)^{-1} d^{-2}}.
 \eeq

Thus, relations (\ref{6.4b})-(\ref{6.6b}) with the moduli function
 \beq{6.10c}
 H = 1 + P R^{-d}
 \eeq
and $P > 0$ from (\ref{6.11b})  give us a charged dilatonic black
hole solution which coincides up to notations with the solutions
from \cite{BBFM,BI}. In \cite{BI} another radial variable  $r =
(R^d + P)^{1/d}$ and parameters $B_{-} = P$ and $B_{+} = P + 2
\mu$ were used. For special cases of this solution see \cite{GM}
($d_0 = 2$), \cite{MP} ($\lambda = 0$, $\varphi$ is absent).

 \subsubsection{Black holes for  $A_2$, $C_2$ and  $G_2$.}

Let us consider the gravitational model with two scalar fields and
two forms of rank 2:
 \beq{6.1}
  S=\int d^Dx \sqrt{|g|} \biggl \{R[g]-
  g^{MN} \partial_M \vec{\varphi} \partial_N \vec{\varphi} -
  \frac{1}{2}   \sum_{s =1}^{2}\exp[2\vec{\lambda_s} \vec{\varphi}](F^s)^2
  \biggr \}.
 \eeq

Here $g$ is a D-dimensional metric, $F^1 = dA^1$ and  $F^2 = dA^2$
are $2$-forms; $\vec{\varphi}=(\varphi^1,\varphi^2) \in \R^2$
 is a vector of two scalar fields,
 $\vec{\lambda_1} = (\lambda_{1 \alpha}), \vec{\lambda_2}
  = (\lambda_{2 \alpha}) \in \R^2$ are dilatonic coupling vectors.

In what follows we  consider  electrically charged black hole
solutions defined on the manifold (\ref{6.2}). We put the brane
multi-indices  corresponding to 1-forms $A^1$ and $A^2$ as $I_1 =
I_2 = \{ 1 \}$, respectively.

We get from (\ref{1a.17})
 \beq{6.3}
  B_{ss'} = \frac{D-3}{D-2} + \vec{\lambda_{s}} \vec{\lambda_{s'}},
 \eeq
 $s, s' = 1,2$, and
  \beq{6.4}
  K_s = B_{ss} = \frac{D-3}{D-2} + \vec{\lambda_{s}}^2 > 0,
  \eeq
$s = 1,2$.

We impose intersection rules (\ref{1a.18}) corresponding to the
Lie algebras $A_2$, $C_2$,  $G_2$ with the Cartan matrices

\beq{6.C.1}
    \left(A_{ss'}\right)=
  \left( \begin{array}{*{6}{c}}
     2 & -1\\
     -k& 2\\
 \end{array}
 \right)\quad
 \eeq
for $k = 1, 2, 3$, respectively. For our case these rules are
equivalent to the set of two relations: (i) $2 B_{12} = A_{12} K_2
= -K_2$, (ii) $2 B_{21} = A_{21} K_1 = - k K_1$, which imply (due
to  $B_{12} = B_{21}$) $K_2 = k K_1$, or
 \beq{6.5}
   \vec{\lambda_{2}}^2 = k \vec{\lambda_{1}}^2 + (k-1) \frac{D-3}{D-2},
  \eeq
  while  relation (ii) reads
 \beq{6.6}
  2  \vec{\lambda_{1}} \vec{\lambda_{2}} = - k \vec{\lambda_{1}}^2 - (k+2) \frac{D-3}{D-2},
  \eeq
 $k = 1, 2, 3$.

The set of relations  (\ref{6.5}), (\ref{6.6}) is equivalent to
intersection rules (\ref{1a.18}).

{\bf Remark.} {\em It may be readily verified that the vectors
 $\vec{\lambda_{1}}, \vec{\lambda_{2}}  \in \R^2$ obeying
(\ref{6.5}), (\ref{6.6}) do exist. This may be done by fixing
 $\vec{\lambda_{1}}^2 = N >0$ and writing the Gramian matrix
 $(\vec{\lambda_{i}} \vec{\lambda_{j}})$ by using (\ref{6.5}) and
(\ref{6.6}) in terms of $N$, $\frac{D-3}{D-2}$ and $k$. For big
enough value of $N$ ($N > N_0$) the Gramian matrix is positive
definite and hence there exist dilatonic coupling vectors which
obey (\ref{6.5}) and (\ref{6.6}).  One may choose:
$\vec{\lambda_{i}} = \sqrt{N} (\vec{u_{i}} + o(1))$, for $N \to +
\infty$, where  $\vec{u_{i}}$, $i = 1,2$, are simple roots of the
Lie algebra  ${\cal G} = A_2, C_2, G_2$ for $k = 1, 2, 3$,
respectively.}

 The charged (by forms $F^1$ and  $F^2$) electric black hole solutions
have the following form (see (\ref{2a.30})-(\ref{2a.32}))

 \bear{6.7}
 g= \Bigl( H_1^{2 h_1}H_2^{2 h_2} \Bigr)^{1/(D-2)}
  \biggl\{ f^{-1} dR \otimes dR + R^2  g^0 \\ \nn
 -   H_1^{-2 h_1}H_2^{-2 h_2}  f  dt \otimes dt +
  g^2  \biggr\},  \\  \label{6.8}
 \exp(\varphi^{\alpha})=  H_1^{h_1 \lambda_{1 \alpha}}
                        H_2^{h_2 \lambda_{2 \alpha}},
 \\  \label{6.9a}
 F^1=   \frac{Q_1}{R^{d_0}}  H_{1}^{- 2} H_{2}  dt \wedge dR
 \\  \label{6.9b}
 F^2=   \frac{Q_2}{R^{d_0}}  H_{2}^{- 2} H_{1}^{k}  dt \wedge dR
 \ear
 where $\alpha = 1, 2$ and $k = 1, 2, 3$ for Lie algebras $A_2$, $C_2$,
$G_2$, respectively.

 Here $f =1 - 2\mu/R^{d}$, $\mu > 0$, $d = d_0 -1$, $h_s = K_s^{-1} >0$,
 $Q_s \neq 0$ ($s =1,2$) are charges, $R > R_0 = (2\mu)^{1/d}$,
 and $g^2$ is a Ricci-flat metric on $M_{2}$.

The moduli functions $H_s$ are given by general prescriptions
(\ref{5.2}) and (\ref{5.3}):

\beq{6.10}
  H_s(z) = \gamma_s^{-1} {\cal H}_{s}(f(z); p), \qquad
  \gamma_s = {\cal H}_{s}(1; p),
\eeq

$s = 1,2$,  where $f(z) = 1 - 2 \mu z$ and the vector $p =
 (p_1, p_2)$ has negative components: $p_1 < 0$ and
  $p_2 < 0$, which are related to  charges $Q_1,
 Q_2$ by formulae

\bear{6.11}
 n_1 p_1 \gamma_1^{-2} \gamma_2
                  =  - K_1 Q_1^2 /(2 \mu d)^2,
                  \\ \label{6.7a}
 n_2 p_2 \gamma_2^{-2} \gamma_1^{k}
                   =  - K_2 Q_2^2 /(2 \mu d)^2,
\ear
$k =1,2,3.$

 The fluxbrane polynomials ${\cal H}_{1},
{\cal H}_{2}$  are given by relations: (i) (\ref{A.6}),
(\ref{A.7}) for $k = 1$; (ii) (\ref{C.2}), (\ref{C.3}) for $k = 2$;
and (iii) (\ref{G.2}), (\ref{G.3}) for $k = 3$. The powers of
polynomials  read: $(n_1,n_2) = (2,2), (3,4),
(6,10)$ for $k = 1, 2, 3$, respectively.

For the Hawking temperatures of black holes  ($k = 1,2,3$) we get from
(\ref{5.7})
 \beq{6.12}
 T_H=   \frac{d}{4 \pi (2 \mu)^{1/d}}
 \prod_{s = 1}^{2} \gamma_{s}^{h_s}. \eeq

We remind that here as in general case the set of parameters $p =
(p_1, p_2)$ should be  proper.

The $A_2$-solution without ``internal space'' $(M_2,g^2)$ and with
one scalar field was obtained recently in \cite{LY}. In \cite{LY}
$A_n$ black hole solutions with $n$ vectors fields and $(n-1)$
scalar fields were found.

\section{Conclusions and discussions}

Here we have described  a family of composite  black brane
solutions corresponding to semisimple Lie algebras in the models
with scalar field and fields of forms. Intersection rules for
branes are given by Cartan matrices for these Lie algebras. The
metric of any solution contains $(n -1)$ Ricci-flat ``internal''
metrics and certain restrictions on brane intersections are
imposed.

The moduli functions of solutions are given by fluxbrane
polynomials which define  special solutions to  open Toda chain
  equations corresponding to semisimple Lie algebras.
 These polynomials may be calculated (in principle) for any
simple or semisimple Lie algebra, e.g. by using MAPLE or
MATHEMATICA. One can   use also  formal relations for $A_m$-
polynomials which were obtained recently in \cite{LY}.
$C_m$-polynomials may be obtained from $A_{2m+1}$-polynomials by
using the identifications of parameters: $p_1 = p_{2m+1}$, $p_2 =
p_{2m}$ etc \cite{IMp3,Iflux}.

Here we have applied our formalism to  few examples of dilatonic
charged black hole ($0$-brane) solutions related to Lie algebras
$A_1$, $A_2$, $C_2$ and  $G_2$. To our knowledge the last solution
is a new one while others are covered by  earlier publications,
e.g.  \cite{BBFM,BI} ($A_1$), \cite{IMp3} ($A_2$) and
\cite{GrIvKim1} ($C_2$). (In the $C_2$-case the parametrization of
polynomials in \cite{GrIvKim1}  differs from the one considered in
this paper).

We have also obtained formulae for  Hawking temperatures
corresponding to  black brane solutions (under consideration) in
terms of polynomial parameters. The calculation of other
thermodynamic quantities and studying the  global structures of
certain black brane/hole solutions will be a subject of a separate
publication.

\begin{center}
 {\bf Acknowledgments}
 \end{center}

This work was supported in part by Templan grant of PFUR (No 200312-1-174) in 2014.



 \end{document}